\documentclass[prl,twocolumn,amsmath,amssymb,groupedaddress,superscriptaddress,floatfix]{revtex4-1}
\usepackage{graphicx}
\usepackage{color}

\begin{document}

\title{Quantum Levy flights and multifractality of dipolar excitations in a random system}

\author{X. \surname{Deng}}
\affiliation{Institut f\"ur Theoretische Physik, Leibniz Universit\"at Hannover, Appelstr. 2, 30167 Hannover, Germany}
\author{B.~L. \surname{Altshuler}}
\affiliation{Physics Department, Columbia University, 538 West 120th Street, New York, New York 10027, USA}
\affiliation{Kavli Institute for Theoretical Physics, University of California, Santa Barbara, CA 93106, USA}
\author{G.~V. \surname{Shlyapnikov}}
\affiliation{LPTMS, CNRS, Univ. Paris-Sud, Universit\'e Paris-Saclay, Orsay 91405, France}
\affiliation{\mbox{Van der Waals-Zeeman Institute, University of Amsterdam, Science Park 904, 1098 XH Amsterdam, The Netherlands}}
\affiliation{Russian Quantum Center, Skolkovo, Moscow Region 143025, Russia}
\affiliation{Wuhan Institute of Physics and Mathematics, Chinese Academy of Sciences, Wuhan 430071, China}
\affiliation{Kavli Institute for Theoretical Physics, University of California, Santa Barbara, CA 93106, USA}
\author{L. \surname{Santos}}
\affiliation{Institut f\"ur Theoretische Physik, Leibniz Universit\"at Hannover, Appelstr. 2, 30167 Hannover, Germany}

\begin{abstract}
We consider dipolar excitations propagating via dipole-induced exchange among immobile molecules randomly spaced in a lattice. 
The character of the propagation is determined by long-range hops (Levy flights).  
We analyze the eigen-energy spectra and the multifractal structure of the wavefunctions. 
In 1D and 2D all states are localized, although in 2D the localization length can be extremely large leading to an effective localization-delocalization crossover in realistic systems. 
In 3D all eigenstates are extended but not always ergodic, and we identify the energy intervals of ergodic and non-ergodic states. The reduction of the lattice filling induces an ergodic to non-ergodic transition, and the excitations are mostly non-ergodic at low filling.
\end{abstract}

\date{\today}

\maketitle



Quantum transport and the spreading of wave packets in disordered media are known to be suppressed by interference. 
This phenomenon - Anderson localization~\cite{Anderson1958},  has been reported in a variety of systems, 
including ultra-sound~\cite{Weaver1990}, microwaves~\cite{Dalichaouch1991}, light~\cite{Wiersma1997},  electrons~\cite{Akkermans2006}, and cold atoms \cite{Billy2008,Roati2008}.
All eigenstates of a quantum particle in disorder are known to be localized in low dimensions~(1D and 2D)~\cite{Abrahams1979}, 
whereas in 3D there is a mobility edge~(ME) separating localized from extended states. 
The extended states are commonly believed to be ergodic: the spatial average of any observable for a given realization of disorder is equivalent to the ensemble average. At the same time the states at the ME  are multifractal~\cite{Wegner1981,Altshuler1986}, i.e. neither localized nor ergodic. 

The ergodicity of many-body wave functions, which is a cornerstone of conventional Statistical Physics, is obviously violated in the regime of many-body localization~\cite{Basko2006}. 
Moreover, there can a finite range of energies, where the states are extended non-ergodic (NEE)~\cite{Pino2015} and present an intriguing multifractal nature. 
One-particle eigenstates of the Anderson model~\cite{Anderson1958} on hierarchical lattices, such as the Bethe lattice~\cite{AbouChacra1973}, are believed to mimic those 
for generic many-body systems. Recent studies of the Bethe lattice~\cite{Biroli2012,DeLuca2014} and random matrix models~\cite{Kravtsov2015} 
have brought additional evidence for the existence of a finite-width band of NEE states. This is in contrast to the ordinary Anderson model, where only states at the ME are multifractal.
 
Being crucial for describing a broad class of systems, the nature of the transition from NEE to extended ergodic (EE) states is far from being understood and even the existence of the NEE phase remains questionable. Therefore, it is interesting to broaden the class of NEE systems that are accessible for theoretical and experimental studies. Below we demonstrate that the eigenstates of a quantum particle 
in a disordered lattice with long-range hops~(quantum Levy flights)~\cite{Levitov1990,Aleiner2011} can be of the NEE type. 
Levy flights, which are difficult to realize experimentally for material particles, appear naturally in the transport of excitations in systems of 
nuclear spins~\cite{Alvarez2015}, nitrogen-vacancy centers~\cite{Neumann2010}, trapped ions~\cite{Richerme2014,Jurcevic2014}, 
Rydberg atoms~\cite{Guenter2013}, and magnetic atoms and polar molecules in optical lattices~\cite{DePaz2013,Yan2013}. 
We focus on the latter system, although our analysis applies to all of them.

Polar molecules in the lowest ro-vibrational state ~\cite{Ni2008,Takehoshi2014,Park2015} can be excited to a second rotational state, building a pseudo-spin-$1/2$ system. In a deep lattice these excitations 
propagate among the (immobile) molecules due to dipole-induced non-radiative excitation transfer.
The hopping amplitude of the excitation from an excited to ground-state molecule decays as $1/r^3$, with $r$ being the intermolecular separation. 
This excitation exchange has been recently observed for KRb molecules~\cite{Yan2013}, opening perspectives for realization 
of spin models~\cite{Micheli2006,Gorshkov2011,Baranov2012}. Typically only a fraction of the lattice is randomly filled by molecules, with maximally one molecule per site~\cite{Yan2013,Moses2015}. 
Thus, exchange of excitations results in a peculiar off-diagonal disorder with long-range hops~\cite{Yao2014}. 

Here we study the spectral statistics and the spectrum of fractal dimensions ~\cite{DeLuca2014} of the eigenstates of a dipolar excitation in this system.   
Due to Levy flights the eigenstates are dramatically different from the eigenstates of a conventional Anderson model and depend crucially on the dimensionality and lattice filling. 
In 1D and 2D lattices all eigenstates are indeed localized. In the latter case, however, the localization length can be extremely large and typical 
experiments encounter finite-size effects ~\cite{Xu2015}. 
In contrast, dipolar excitations in 3D systems are always extended. The spectrum contains however both NEE and EE regions. The former grows when the lattice filling decreases, and hence 
3D dipolar excitations are dominantly non-ergodic at low fillings.



\paragraph{Model.--} In the following we consider molecules in two possible rotational states, the lowest ro-vibrational state~($\downarrow$) and an excited one~($\uparrow$). 
The molecules are confined in a cubic lattice (square lattice for 2D systems) in the absence of any external electric field.
We will assume for simplicity that all molecules are $\downarrow$, and there is a single excitation corresponding to a spin flip. Similar results are expected for multiple excitations as long as the gas of excitations remains sufficiently dilute. 
Due to dipole-dipole interactions, the excitation may be transferred from a molecule $i$ to another molecule $j$, with a hopping amplitude:
\begin{equation}
t_{ij}=\frac{-d_{\uparrow\downarrow}^2}{a^3 |{\mathbf r}_i-{\mathbf r}_j|^3}
\left (1-3\cos^2\theta_{ij} \right ) 
\end{equation}
where $d_{\uparrow\downarrow}$ is the dipole matrix element between $\downarrow$ and $\uparrow$ states, ${\mathbf r}_j$ is the position of the $j$-th molecule in units of the lattice spacing $a$, and $\theta_{ij}$ is the angle between the quantization axis and the vector $({\bf r}_i-{\bf r}_j)$. The motion of the excitation can then be described by an effective single-particle Hamiltionian with long-range anisotropic hopping:
\begin{eqnarray}
H=-\sum_{i,j}t_{ij}|i\rangle\langle j|,
\label{eq:H}
\end{eqnarray}
where $|j\rangle$ denotes the state in which the excitation is at molecule $j$. We assume that the molecule positions ${\mathbf r}_j$ are randomly distributed with an average 
filling factor $0\leq \rho \leq 1$, and accordingly the couplings $t_{ij}$ are also random. Below we measure all energies in units of $t_0=-d_{\uparrow\downarrow}^2/a^3$, and all lengths in units of the lattice constant $a$. Effective dimensionless parameter of the problem is the filling factor $\rho$, and the disorder is maximal in the limit of $\rho\rightarrow 0$. For growing $\rho$, the hopping terms $t_{ij}$ become more regular, and at $\rho=1$  
the dipolar excitations propagate ballistically in a regular lattice.

We study the properties of model~\eqref{eq:H} by means of exact diagonalization for different dimensionalities $d$ and fillings $\rho$. We consider from $100$ to $1000$ 
random realizations for each $d$ and $\rho$, 
characterized by a random distribution of $N$ molecules in  $L^d$ lattice sites, with $\rho=N/L^d$. Our numerical capabilities limit the number of molecules to $N=80000$. The results from finite-size systems are then extrapolated to correctly infer the asymptotic properties for $N\rightarrow\infty$.


\begin{figure}  [t]
\includegraphics[width=1\linewidth]{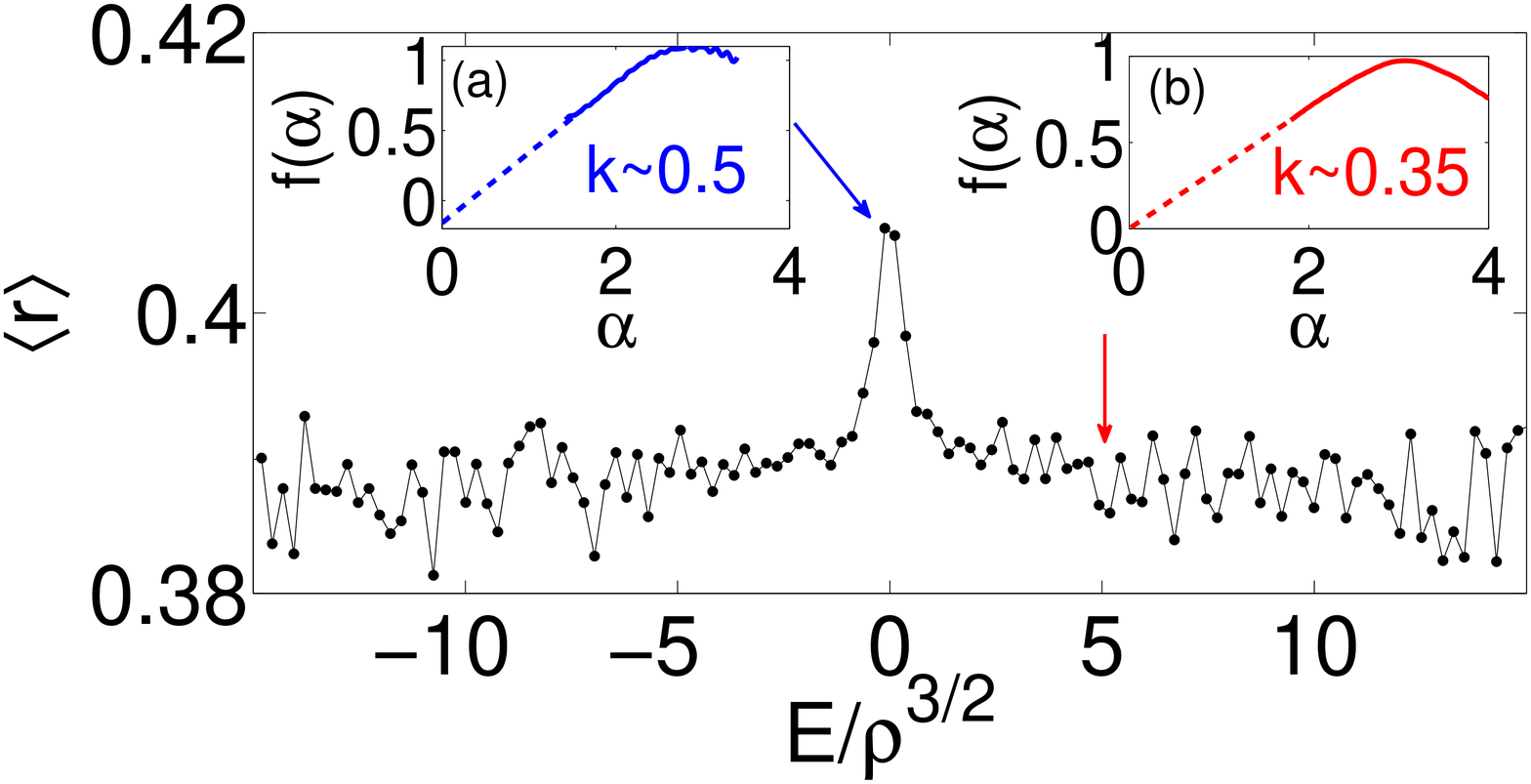}
\caption{(Color online) Dilute limit in 2D: $\langle r \rangle$ as a function of $E$ for $L=9\times10^4$ and $\rho=2.5\times 10^{-6}$, averaged over $1000$ samples. 
The right inset shows the triangular $f(\alpha)$ with slope $k\approx 0.35$ at $E\rho^{-3/2}=5$, whereas 
the left inset shows the case of $E\rho^{-3/2}=0.5$ where $k\simeq 0.5$.}
\label{2D_dilute}
\end{figure}



\paragraph{Eigenstate properties.--} We focus below on the properties of the eigenstates of  Hamiltonian~\eqref{eq:H}, $H |\psi_n\rangle=E_n|\psi_n\rangle$. 
The distribution of the level spacings, $\delta_n=E_{n+1}-E_{n}$, is best characterized by the ratio ~\cite{Shklovskii1993,Huse2007}
 
\begin{equation}   \label{rn}
r_n=\min{(\delta_n,\delta_{n-1})}/\max{(\delta_n,\delta_{n-1})}.
\end{equation}
Localized states present a Poissonian distribution of $r_n$, with an average $\langle r \rangle_{P}\approx0.386$ due to the proliferation of 
degenerate states located at distant spatial positions. In contrast, extended states show level repulsion, displaying a Gaussian orthogonal ensemble~(GOE) Wigner-Dyson distribution of $r_n$, 
with $\langle r\rangle_{GOE}\approx0.53$. We evaluate $\langle r\rangle$ in different energy windows.

The spatial properties of the eigenfunctions, $|\psi_n\rangle=\sum_j \psi_n(j) |j \rangle$, are best characterized by the moments 
\begin{equation}     \label{Iq}
I_q(n)=\sum_j|\psi_n(j)|^{2q}\propto N^{-\tau_q(n)}. 
\end{equation}
The inverse participation ratio~(IPR) $I_2$ measures the inverse of the number of molecules participating in a given eigenstate. 
For extended ergodic~(EE) states the ensemble average $\langle |\psi_n(j)|^{2q} \rangle$ matches the spatial average. The EE states 
present $\tau_q=(q-1)$, whereas $\tau_q=0$ for localized states~\cite{Wegner1981,DeLuca2014}. 

In general, it is convenient to introduce fractal dimensions 
\begin{equation}      \label{Dq}
D_q=\tau_q/(q-1),
\end{equation}
so that for EE states we have $D_q=1$, whereas $D_q=0$ for localized states. Non-ergodic extended (NEE) states are multifractal ~\cite{Wegner1981,Altshuler1986}, i.e. $0<D_q<1$ and  
$D_q$ decreases with increasing $q$.
Following the procedure of Ref.~\cite{DeLuca2014} we perform the Legendre transform of $\tau_q$ and evaluate the so-called spectrum of fractal dimensions~(SFD) $f(\alpha)$. 
This function characterizes the Hausdorff dimension of the sites 
with a probability density $|\psi_j|^2=N^{-\alpha}$. For EE states, $f(\alpha)$ shows a delta-functional behavior, namely $f=1$ at $\alpha=1$ and is equal to $-\infty$ otherwise. For NEE states $f(\alpha)$ presents 
a parabolic form with an exact symmetry $f(1+x)=f(1-x)+x$~\cite{AP,MF}. Localized states display a triangular form, and $f(\alpha)=k\alpha$ in the left part of the triangle, with $k<1/2$~($k=1/2$ characterizes critical states with diverging localization length).

The eigenstate properties depend crucially on dimensionality. For 1D systems excitations remain localized for any  $\rho<1$. 
We have checked in particular that for $\rho=0.99$ the distribution of level spacings, $r_n$, remains Poissonian, and $D_2=0$ in the whole spectrum. 
Localized states (also in 2D) present exponential localization at intermediate distances and wings following a $1/r^6$ decay, resulting from the $1/r^3$ dependence of long-range hops.


\paragraph{2D Systems.-- } Let the molecules be located in the $xy$ plane. The quantization axis is in the $xz$ plane and it forms an angle $\beta$ with the $x$ direction. We thus have $t_{ij}=t_0 (1-3\cos^2\beta\cos^2\phi)/|{\mathbf r}_i-{\mathbf r}_j|^3$, with $\phi$ being the angle between $({\mathbf r}_i-{\mathbf r}_j)$ and the $x$ direction. We assume $\cos^2\beta=2/3$, in order to mimic best the 3D anisotropy. However, in the isotropic case, $\beta=\pi/2$, the results are quite similar. 

We start the description of our results with the dilute limit $\rho\ll 1$, where the intermolecular spacing is much larger than the lattice spacing and the properties of the system are self-similar with the filling $\rho$. The relevant energy scale is $\rho^{3/2}$, corresponding to the dipole-dipole interaction at the mean intermolecular distance. 
Figure~\ref{2D_dilute} shows $\langle r \rangle$ for different energy windows. 
Eigenstates with energies $|E|\gtrsim \rho^{3/2}$ are clearly localized, with $\langle r \rangle\simeq 0.386$ and a triangular SFD with slope $k<1/2$~(see the right inset to Fig.~\ref{2D_dilute}). 
In contrast, for $|E|\rho^{-3/2}< 1$ we have $\langle r \rangle>0.386$, and the SFD $f(\alpha)$ is also triangular but with the slope $k\simeq 1/2$~(left inset of Fig.~\ref{2D_dilute}). This indicates  
that the inner states are either critical or have an extremely large localization length.

\begin{figure}
\includegraphics[width=1.0\linewidth]{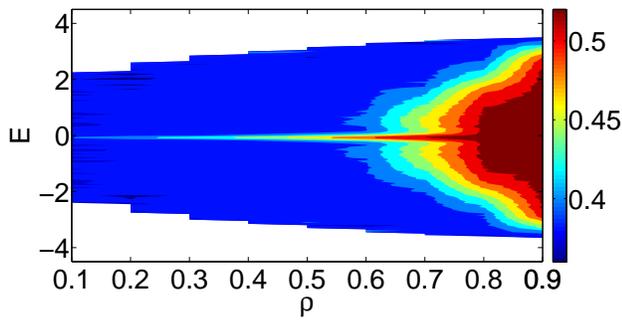}
\caption{(Color online) Spectral statistics at finite filling in 2D: $\langle{r}\rangle$ in various energy windows for lattices with fillings $\rho$.}
\label{2D_r_rho}
\end{figure}

We then consider 2D systems at a finite filling.  Figure~ \ref{2D_r_rho} shows the distribution of  $\langle r\rangle$ for $N=50000$ and different values of $\rho$. 
For $\rho \leq 0.5$ the distribution of $\langle r \rangle$ closely resembles that in the dilute limit, with a central critical core of low-$|E|$ states, and localized states outside of this core. 
For $\rho\geq 0.6$ a growing central region of the spectrum presents $\langle r \rangle > 0.386$. The effective level repulsion results from the finite size of the system, 
since the localization length of these states is likely comparable to the system size, hence driving a localization to delocalization~(L-DL) crossover ~\cite{note2}. 



\begin{figure}
\includegraphics[width=1.0\linewidth]{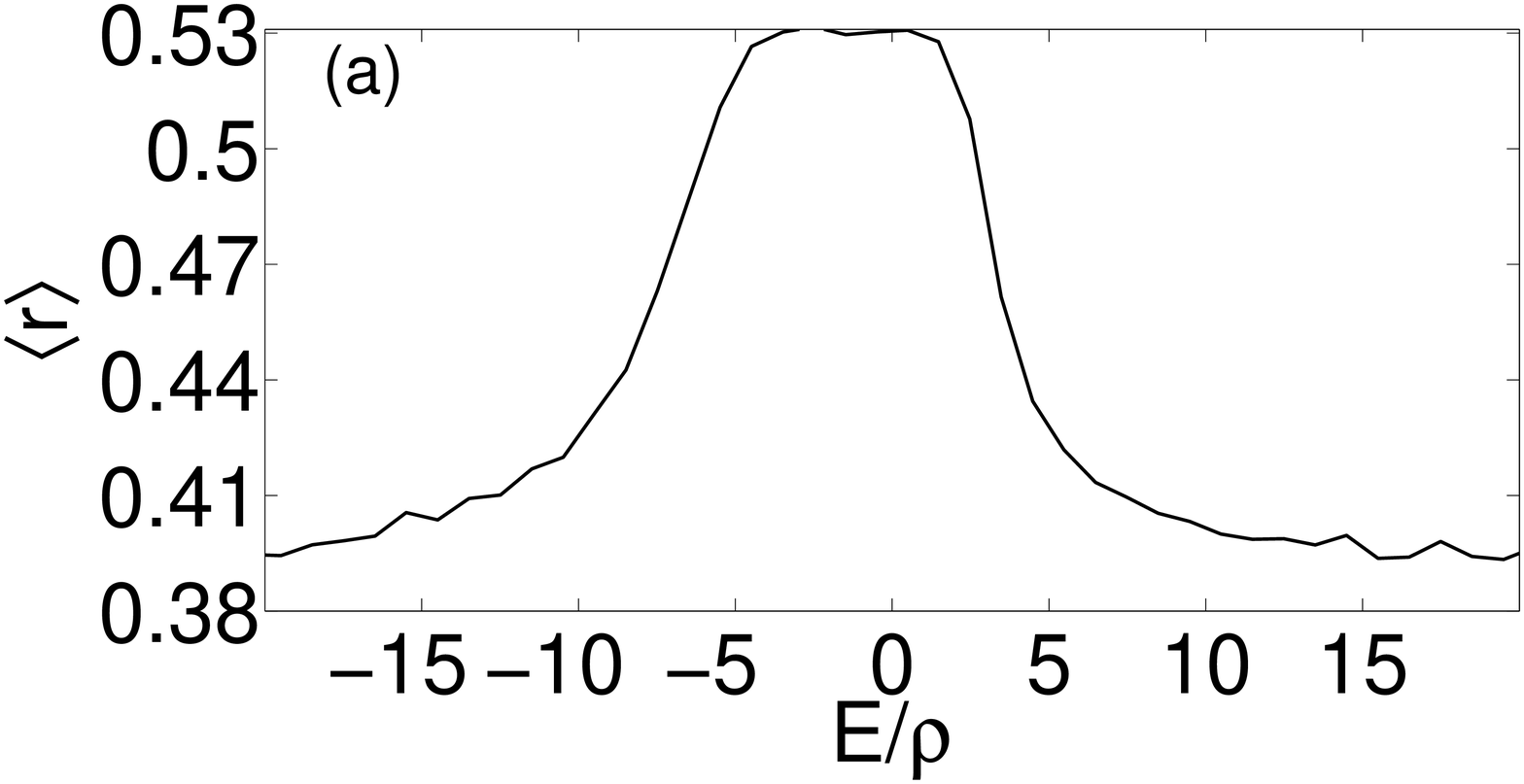}
\includegraphics[width=1.0\linewidth]{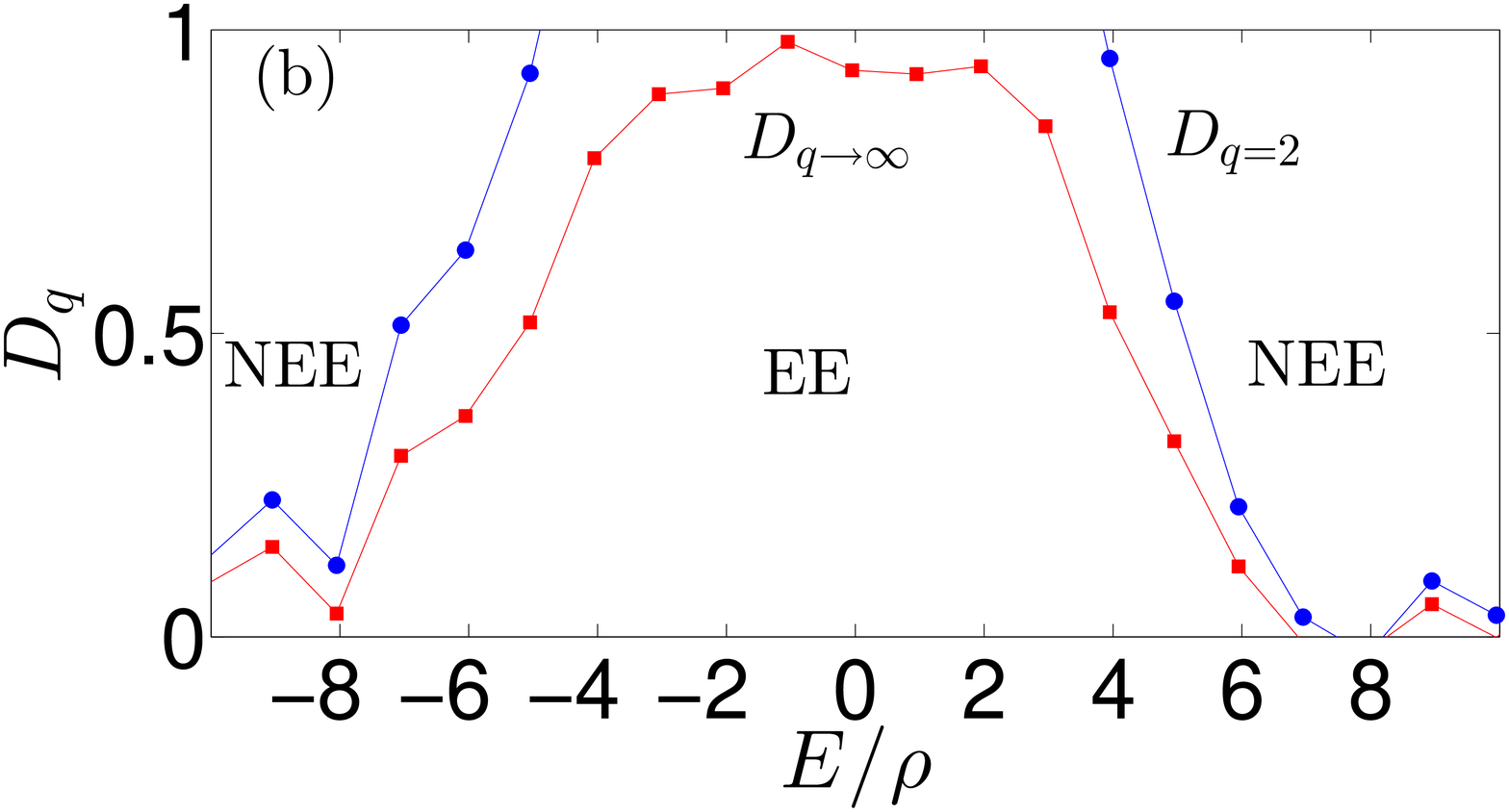}
\caption{(Color online) Dilute limit in 3D: (a) $\langle r\rangle$ as a function of $E/\rho$ for $L=1800$ and $\rho=10^{-5}$, averaged over $100$ disorder samples. 
(b) $D_{q=2}$ and $D_{q\to\infty}$ as functions of energy.}
\label{3D_dilute}
\end{figure}



\begin{figure} [t]
\includegraphics[width=1.\linewidth]{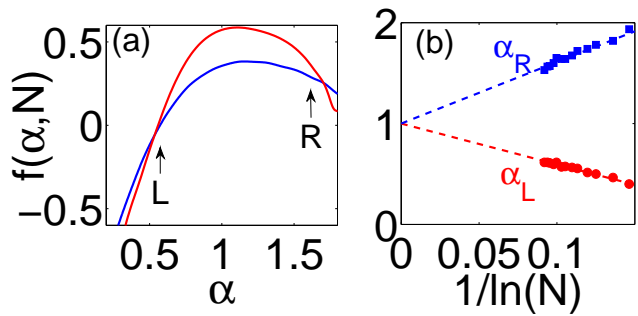}
\caption{(Color online) SFD for $\rho=0.1$: (a) $f(\alpha,N)$ at $E=0$ for $N=1500$~(blue curve) and $N=61000$~(red curve). Note the crossing points $L$ and $R$ at $\alpha_{R,L}$,  
between the SFD for the two values of $N$. 
(b) The crossings $\alpha_{R,L}$ follow a $1/\ln N$ extrapolation all the way to $1$, as expected for EE states.
}
\label{3D_dilute_crossings}
\end{figure}

\paragraph{3D Systems.--} Figures~\ref{3D_dilute}(a) and~(b) summarize our 3D results 
in the dilute limit for $\langle r\rangle$, $D_2$ and $D_{\infty}$~\cite{footnote-Dq}. Note that in 3D the characteristic energy scale, given by the dipolar interaction at the 
mean interparticle distance is $\rho$. 
Eigenstates with $|E|\rho^{-1}\lesssim 2$ are EE, characterized by 
$\langle r \rangle\simeq 0.53$ and $D_q\simeq 1$ for all values of $q$. The EE character is confirmed by our analysis of the SFD for 
growing system sizes $\{ N_1<N_2<\dots < N_s < \dots  \}$. Motivated by Ref.~\cite{DeLuca2014}, we evaluate the crossings between $f(\alpha,N_{s})$ and $f(\alpha,N_{s+1})$~(Fig.~\ref{3D_dilute_crossings}(a)).  
We recall that NEE states display a parabolic SFD, whereas for EE states $f(\alpha=1)=1$ and $f(\alpha\neq 1)=-\infty$. 
For $E=0$ the crossings converge towards $\alpha=1$ following a $1/\ln N$ dependence~(Fig.~\ref{3D_dilute_crossings}(b)). Hence, in the center of the band the states are extended ergodic.

\begin{figure} [t]
\includegraphics[width=1.\linewidth]{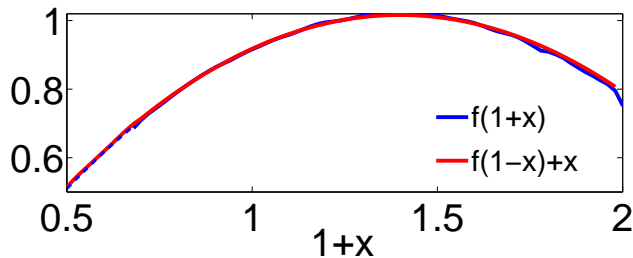}
\caption{(Color online) SFD at $E=-1$ for $\rho=0.1$, presenting a parabolic form that fulfills the symmetry $f(1+x)=f(1-x)+x$, characteristic of NEE states.}
\label{3D_dilute_crossings_2}
\end{figure}

For eigenstates outside the center we have $0.386<\langle r \rangle < 0.53$, and $0<D_q<1$ that decay with growing $q$. 
The SFD is parabolic and fulfills  the relation $f(1+x)=f(1-x)+x$~(Fig.~\ref{3D_dilute_crossings_2}). Hence, these are clearly NEE states. Thus, for $\rho\ll 1$, moving from the center to the wings of the spectrum we have a transition from EE to NEE states. There is no Anderson transition into the localized regime even at the spectral wings. The eigenstates remain NEE, 
with $f(\alpha)$ approaching the critical triangular form with slope $1/2$ as $|E|$ increases.

\begin{figure}[t]
\includegraphics[width=1.0\linewidth]{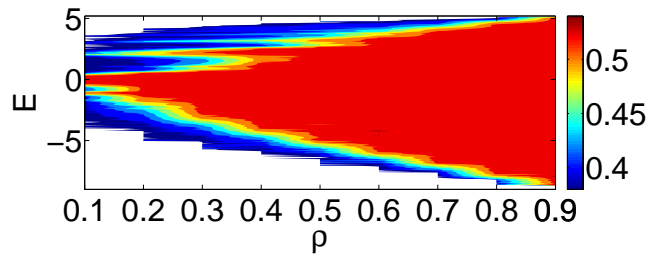}
\vspace*{-0.4cm}
\caption{(Color online) Spectral statistics at finite filling in 3D: $\langle{r}\rangle$ in various energy windows for lattices with fillings $\rho$.}
\label{3D_aniso}
\end{figure}

The ergodic to non-ergodic transition is also observed at finite $\rho$. The central region, where the EE character of the states is confirmed by the SFD crossing technique discussed above, 
broadens towards eigenstates of large $|E|$ for growing $\rho$ and pushes the NEE region to the spectral outscores. This behavior 
is illustrated by the $E$-dependence of $\langle r \rangle $ in Fig.~\ref{3D_aniso} for different values of $\rho$. Note that the central EE region with $\langle r \rangle\simeq 0.53$ 
broadens and covers basically all the spectrum already for $\rho\simeq 0.5$. A small sliver of NEE states remains at the spectral borders.



\paragraph{Conclusions and outlook.--}
Excitations propagating via dipole-induced exchange among randomly distributed particles in a lattice conform a peculiar effectively disordered system, whose properties 
depend crucially on the lattice filling $\rho$ and on dimensionality. 
One-dimensional systems are clearly localized all the way up to $\rho=1$. In 2D all eigenstates are localized at $\rho<1$, with a very large or even diverging localization length in the middle of the band. At $\rho$ close to half-filling, finite 2D systems under typical experimental conditions should experience an effective L-DL crossover. 

In 3D all eigenstates are extended. The states are EE in the center of the band, and outside the center they are NEE. 
A change of the filling factor $\rho$ can induce an ergodic$\leftrightarrow$non-ergodic transition for a fixed energy $E$. The NEE spectral 
region exists at any filling factor, and it becomes dominant for small $\rho$.

Currently it is possible to realize experimentally a lattice with $\sim 10^6$ sites and filling factor up to $\rho \sim 0.3$ or even higher. One can think of the following experiment: create a dipolar excitation in a particular site of the lattice and measure the probability $P(r,t)$ to find it at a distance r after time t. In the fully ergodic case this probability has a familiar diffusion distribution, with the diffusion constant $D$:
\begin{equation}      
P(r,t)=(Dt)^{-3/2}\exp(-r^2/Dt).
\label{eq:P}
\end{equation}					
The broadening of the wave packet in the non-ergodic case is much slower. The initial state involves both ergodic and non-ergodic eigenstates. It is safe to expect that $P(r,t)$ is determined by the ergodic states and thus follows the law~\eqref{eq:P} at large distances exceeding $\sim\sqrt{Dt}$. However, at smaller distances $P(r,t)$ is determined by NEE states and thus substantially exceeds the result of Eq.~\eqref{eq:P}. In particular, the return probability $P(r=0,t)$ has a nontrivial power-like time dependence. A detailed analysis of dynamical properties of dipolar excitations will be presented elsewhere. 


\acknowledgments
We are grateful to V.E. Kravtsov for fruitful discussions. 
We acknowledge support by the Center QUEST and the DFG Research Training Group 1729, and the support from IFRAF and the Dutch foundation FOM. The research leading to these results has received funding from the European Research Council under European Community's Seventh Framework Programme (FR7/2007-2013 Grant Agreement no. 341197). This research was supported in part by the National Science Foundation under Grant No. NSF PHY11-25915.



\begin{thebibliography}{99}
\bibitem{footnote-Dq} We calculate $D_q$ directly up to $q=10$ for growing system sizes $N_s$. In order to get $D_q$ at $N\rightarrow\infty$ 
we extrapolate $D(q,N)$ using the relation $D(q,N)=k_q/\ln(N)+D_q$, where $N=(N_s+N_{s+1})/2$, and $k_q$ is a fitting coefficient. 
The value $D_{q\to\infty}$ is obtained from a linear extrapolation of $D_q$ as a function of $1/q$.
\end{thebibliography}
\end{document}